%
\input harvmac
\def\Title#1#2#3{#3\hfill\break \vskip -0.35in
\rightline{#1}\ifx\answ\bigans\nopagenumbers\pageno0\vskip.2in
\else\pageno1\vskip.2in\fi \centerline{\titlefont #2}\vskip .1in}

\font\ticp=cmcsc10
\def\ajou#1&#2(#3){\ \sl#1\bf#2\rm(19#3)}
\Title{}{Night Thoughts of a Quantum Physicist}{~}

\centerline{{\ticp Adrian Kent}}
\vskip.1in
\centerline{\sl Department of Applied Mathematics and
Theoretical Physics,}
\centerline{\sl University of Cambridge,}
\centerline{\sl Silver Street, Cambridge CB3 9EW, U.K.}

\bigskip
\centerline{\bf Abstract} 

The most dramatic developments in theoretical physics in the next
millennium are likely to come when we make progress on so far
unresolved foundational questions.  In this essay I consider two
of the deepest problems confronting us, the measurement problem in 
quantum theory and the problem of relating consciousness to the
rest of physics.  I survey some recent promising ideas on 
possible solutions to the measurement problem and explain what 
a proper physical understanding of consciousness would involve
and why it would need new physics.  

\medskip\noindent
\newsec{Introduction} 

As the twentieth century draws to a close, theoretical physics
is in a situation that, at least in recent history, is most
unusual: there is no generally accepted authority.  
Each research program has very widely respected leaders, 
but every program is controversial.   
After a period of extraordinary successes, broadly stretching from 
the 1900's through to the early 1980's, there have been
few dramatic new experimental results in the last fifteen
years, with the important exception of cosmology.  
All the most interesting theoretical ideas 
have run into serious difficulties, and it is not completely obvious
that any of them is heading in the right direction.  
So to speak, some impressively large and well organised 
expeditionary parties have been formed 
and are faithfully heading towards imagined
destinations; other smaller and less cohesive bands
of physicists are heading in quite different directions.  
But we really are all in the dark.  Possibly none of us 
will get anywhere much until the next fortuitous break in the clouds.

I will try to sketch briefly how it is that we have reached this 
state, and then suggest some new 
directions in which progress may eventually be possible.
But my first duty is to stress 
that what follow are simply my personal views.  
These lie somewhere between the heretical and the mainstream at the 
moment. 
Some of the best physicists of the twentieth century, 
would, I think, have been at least in partial 
sympathy.\foot{In any case, I am greatly indebted to 
Schr\"odinger and Bell's lucid scepticism 
and to Feynman's compelling explanations of the scientific need 
to keep alternative ideas in mind if they are even partially
successful, as expressed in, for example, Schr\"odinger 1954, Bell 1987,
Feynman 1965.}
But most leading present day
physicsts would emphasize different problems;
some would query whether physicists can sensibly
say anything at all on the topics I will discuss.

I think we can, of course.  It seems to me the problems are  
as sharply defined as those we have overcome in the past: it just 
happens that we have not properly tackled them yet.  They would be 
quite untouched --- would remain deep
unsolved problems --- even if what is usually meant by
a ``theory of everything'' were discovered.
Solving them may need further radical
changes in our world view, but I suspect that in the end 
we will find there is no way around them.  
 
\newsec{Physics in 1999}

The great discoveries of twentieth century physics have sunk
so deeply into the general consciousness that it now
takes an effort of will to stand back and try to see 
them afresh. 
But we should try, just as we should try to look at the night sky and at 
life on earth with childlike eyes from time to time.  In 
appreciating just how completely and how 
amazingly our understanding of the world has been transformed,
we recapture a sense of awe and wonder in the universe and its beauty.\foot{
We owe this, of course, not to nature --- which gives a very
good impression of not caring either way --- but to ourselves.  
Though we forget it too easily, that sense is precious to us.} 

So recall: 
in 1900, the existence of atoms was a controversial
hypothesis.  Matter and light were, as far as we knew, qualitatively
different.  The known laws of nature were deterministic and 
relied on absolute notions of space and time which seemed 
not only natural and common sense but also so firmly embedded
in our understanding of nature as to be beyond serious question.
The propagation of life, and the functioning of the mind, 
remained so mysterious that it was easy to imagine their
understanding might require quite new physical principles. 
Nothing much resembling modern cosmology existed.  

Einstein, of course, taught us to see space and time as
different facets of a single geometry.
And then, 
still more astonishingly and beautifully, that the 
geometry of spacetime is nonlinear, that matter 
is guided by the geometry and at the same time shapes it, 
so that gravity is understood as the mutual action
of matter on matter through the curvature of spacetime. 

The first experiments confirming 
an important prediction of general relativity ---
that light is indeed deflected by the solar gravitational field ---
took place in 1917: still within living memory.  
Subsequent experimental tests have confirmed
general relativity with increasingly impressive accuracy.  
It is consistent with our understanding of cosmology,  
as far as it can be --- that is, as far as quantum effects
are negligible.  At the moment it has no remotely serious
competitor: we have no other picture of
the macroscopic world that makes sense and fits the data.  

Had theorists been more timid, particle physics experiments and
astronomical observations would almost certainly eventually 
given us enough clues to make the development of special and 
general relativity inevitable.  
As it happens, though, Einstein was only partially  
guided by experiment.   
The development of the theories of relativity relied
on his extraordinary genius for seeing through to  
new conceptual frameworks underlying known physics. 
To Einstein and many of his contemporaries, 
the gain in elegance and simplicity was so great
that it seemed the new theories almost had to be correct.  

While the development of quantum theory too relied on brilliant
intuitions and syntheses, it was much more driven by experiment.
Data --- the black-body radiation spectrum, the photo-electric effect, 
crystalline diffraction, atomic spectra --- more or less 
forced the new theory on us, first in ad hoc forms, and then, by 
1926, synthesised.  
It seems unlikely that anyone would ever have
found their way through to quantum theory unaided by the
data.  Certainly, no one has ever found a convincing conceptual 
framework which explains to 
us why something like quantum theory should be true.  It just is.  
Nor has anyone, even after the event, come up with a truly
satisfactory explanation of what precisely
quantum theory tells us about nature.  We know that all
our pre-1900 intuitions, based as they are on the physics
of the world we see around us every day, are quite inadequate.
We know that microscopic systems behave in a qualitatively
different way, that there is apparently an intrinsic randomness in the way
they interact with the devices we use to probe them.
Much more impressively, for any given experiment
we carry out on microscopic systems, we know how to 
list the possible outcomes and calculate the probabilities 
of each, at least to a very good approximation.  
What we do not fully understand is why those calculations 
work: we have, for example, no firmly established 
picture of what (if anything) 
is going on when we are not looking. 

Quantum theory as originally formulated was inconsistent 
with special relativity.  Partly for this reason, it did not properly
describe the interactions between light and matter either.
Solving these problems took several further steps, and in time 
led to a relatively systematic
--- though still today incomplete --- 
understanding of how to build relativistic quantum theories
of fields, and eventually to the conclusion that
the electromagnetic force and the two nuclear forces
could be combined into a single field theory. 
As yet, though, we do not know how to do that very 
elegantly, and almost everyone suspects that 
a grander and more elegant unified theory of those
three forces awaits us.  Nor can we truly say that we fully
understand quantum field theory, or even that the theories
we use are entirely internally consistent.  They resemble
recipes for calculation, together with only partial, though tantalisingly 
suggestive, explanations as to why they work. 
Most theorists believe a deeper 
explanation requires a better theory, perhaps yet to be
discovered.  

Superstring theory, which many physicists hope might
provide a complete theory of gravity as well as the other 
forces--- a ``theory of everything'' --- is currently the most 
popular candidate.  Though no one doubts its mathematical
beauty, it is generally agreed that so far superstring theory 
has two rather serious problems.  
Conceptually, we do not know how to properly
make sense of superstrings as a theory of matter plus 
spacetime.  Nor can we extract any very interesting correct
predictions from the theory --- for example, the properties of
the known forces, the masses of the known particles, 
or the apparent four-dimensionality of space-time ---
in any convincing way.  

Opinions differ sharply on whether those 
problems are likely to be resolved, and so whether
superstring theory is likelier to be a theory of
everything or of nothing: time will tell.
Almost everyone agrees, though, that 
reconciling gravity and quantum theory is one of the deepest
problems facing modern physics.  
Quantum theory and general relativity, each brilliantly 
successful in its own domain, rest on very different principles 
and give highly divergent pictures of nature.  
According to general relativity, the world
is deterministic, the fundamental equations of nature are non-linear,
and the correct picture of nature is, at bottom, geometric. 
According to quantum theory, there is an intrinsic randomness
in nature, its fundamental equations are linear, and the 
correct language in which to describe nature seems to be 
closer to abstract algebra than geometry.  
Something has to give somewhere, but at the moment we do not
know for sure where to begin in trying to combine these pictures: we do
not know how to alter either in the direction of the other
without breaking it totally.  

However, I would like here to try to look a bit
beyond the current conventional wisdom.  
There is always a danger that attention clusters around some
admittedly deep problems while neglecting others, 
simply through convention, or habit or sheer 
comfort in numbers.  Like any other subject, theoretical physics
is quite capable of forming intellectual taboos: topics
that almost all sensible people avoid.  
They often have good reason, of course, but I suspect that the most strongly 
held taboos sometimes resemble a sort of 
unconscious tribute.  Mental blocks can form because a question 
carries the potential for revolution, and
addressing it thoughtfully would raise the possibility
that our present understanding may, in important
ways, be quite inadequate: in other words, they can be unconscious
defences against too great a sense of insecurity.
Just possibly, our best hope of 
saying something about future revolutions in physics may lie 
in looking into interesting questions which current theory evades.  
I will look at two here: the measurement problem in quantum
theory and the mind-body problem.  

\newsec{Quantum Theory and the Measurement Problem} 

As we have already seen, quantum theory was not originally inspired 
by some parsimonious set of principles applied to sparse data.
Physicists were led to it, often without seeing a
clear way ahead, in stages and by a variety of accumulating data.  
The founders of quantum theory were thus immediately faced with
the problem of explaining precisely what the theory
actually tells us about nature.  On this they were never able to agree.
However, an effective enough consensus, led by Bohr, was forged.
Precisely what Bohr actually believed, and why, 
remain obscure to many commentators, but for most practical
purposes it has hardly mattered.  Physicists found 
that they could condense Bohr's ``Copenhagen interpretation'' 
into a few working rules which explain what can usefully
be calculated.  Alongside these, a sort of working 
metaphysical picture --- if that is not a contradiction in 
terms --- also emerged.  
C.P. Snow captures this conventional wisdom well in 
his semi-autobiographical novel, ``The Search'' (Snow 1934):

{\narrower\smallskip\noindent
Suddenly, I heard one of the greatest mathematical physicists say,
with complete simplicity: ``Of course, the fundamental laws of 
physics and chemistry are laid down for ever.  The details have
got to be filled up: we don't know anything of the nucleus; but
the fundamental laws are there.  In a sense, physics and chemistry
are finished sciences.''

The nucleus and life: those were the harder problems: in everything
else, in the whole of chemistry and physics, {\it we were in sight
of the end}.  The framework was laid down; they had put the boundaries
round the pebbles which we could pick up. 

It struck me how impossible it would have been to say this a few years
before.  Before 1926 no one could have said it, unless he were a
megalomaniac or knew no science.  And now two years later the most
detached scientific figure of our time announced it casually in the
course of conversation.  

It is rather difficult to put the importance of this revolution into
words. [$\ldots$] However, it is something like this. Science starts with
facts chosen from the external world. The relation between the choice,
the chooser, the external world and the fact produced is a complicated
one [$\ldots$] but one gets through in the end [$\ldots$] to
an agreement upon
``scientific facts''.  You can call them ``pointer-readings'' as
Eddington does, if you like.  They are lines on a photographic plate,
marks on a screen, all the ``pointer-readings'' which are the end of
the skill, precautions, inventions, of the laboratory.  They are the
end of the manual process, the beginning of the scientific.  For from
these ``pointer-readings'', these scientific facts, the process of 
scientific reasoning begins: and it comes back to 
them to prove itself right or wrong. For the scientific process is
nothing
more nor less than a hiatus between ``pointer-readings'': one takes
some pointer readings,  makes a mental construction from them in
order to predict some more.  

The pointer readings which have been predicted are then measured:
and if the prediction turns out to be right, the mental construction
is, for the moment, a good one.  If it is wrong, another mental
construction has to be tried.  That is all.  And you take your
choice where you put the word ``reality'': you can find your total
reality either in the pointer readings or in the mental construction
or, if you have a taste for compromise, in a mixture of both.\smallskip}  

In other words, on this conventional view, quantum 
theory teaches us something deep and revolutionary about the
nature of reality.  It teaches us that it is a mistake to try to
build a picture of the world which includes every aspect of an
experiment --- the preparation of the apparatus and the
system being experimented on, their behaviour during the 
experiment, and the observation of the results --- in one
smooth and coherent description.  All we need to do science, 
and all we can apparently manage, is to find a way of 
extrapolating predictions --- which as it happens turn out generally to be
probabilistic rather than deterministic --- about the final results from
a description of the initial preparation.  To ask what 
went on in between is, by definition, to ask about something
we did not observe: it is to ask in the abstract a question 
which we have not asked nature in the concrete.  On the
Copenhagen view, it is a profound feature of our situation 
to the world that we cannot separate the abstract and the
concrete in this way.  If we did not actually carry out
the relevant observation, we did not ask the question in
the only way that causes nature to supply an answer, and 
there need not be any meaningful answer at all.

We are in sight of the end.  Quantum theory teaches us the 
necessary limits of science.  But are we?  Does it? 
Need quantum theory be understood only as a
mere device for extrapolating pointer-readings from
pointer-readings?  {\it Can}
quantum theory be satisfactorily understood this way?
After all, as we understand it, a pointer is no more than a 
collection of atoms following quantum laws.  If the atoms and 
the quantum laws are ultimately just mental constructions, is 
not the pointer too?  Is not everything?

Landau and Lifshitz, giving a precise and apparently not 
intentionally critical description of the orthodox view in
their classic textbook (Landau \& Lifshitz, 1974) on 
quantum theory, still seem to hint at some disquiet here: 

{\narrower\smallskip\noindent Quantum mechanics occupies a very
unusual place among physical theories: it contains classical mechanics
as a limiting case, yet at the same time requires this limiting case
for its own formulation. \smallskip}

This is the difficulty.
The classical world --- the world of the laboratory --- must
be external to the theory for us to 
make sense of it; yet it is also supposed to be contained
within the theory.  And, since the same objects play this
dual role, we have no clear division between the
microscopic quantum and the macroscopic classical.   
It follows that we cannot legitimately derive
from quantum theory the predictions we believe the theory
actually makes.  If a pointer is only a mental construction,
we cannot meaningfully ask what state is in or where it
points, and so we cannot make meaningful predictions about
its behaviour at the end of an experiment.  If it is a real object 
independent of the quantum realm, then we cannot explain it --- or, 
presumably, the rest of the macroscopic world around us --- in 
terms of quantum theory.  Either way, if the Copenhagen interpretation
is right, a crucial component in our understanding of the
world cannot be theoretically justified.

However, we now know that Bohr, the Copenhagen
school, and most of the pioneers of quantum theory were unnecessarily
dogmatic.  We are not forced to adopt the Copenhagen interpretation
either by the mathematics of quantum theory or by empirical evidence.
Nor is it the only serious possibility available.
As we now understand, it is just one of several
possible views of quantum theory, each of which has advantages
and difficulties.  It has not yet been superseded: there is 
no clear consensus now as to which view is correct.  But it 
seems unlikely it will ever again be generally accepted as the 
one true orthodoxy.  

What are the alternatives?  The most interesting, I think,
is a simple yet potentially revolutionary idea originally
set out by Ghirardi, Rimini, and Weber (Ghirardi et al. 1986), and later
developed further by GRW, Pearle, Gisin and several others. 
According to their model, quantum mechanics has a piece missing. 
We can fix all its problems by adding rules 
to say exactly how and when the quantum dice are rolled.  
This is done by taking wave function collapse to 
be an objective, observer-independent phenomenon, with small 
localisations or ``mini-collapses'' constantly taking place.
This entails altering the dynamics by adding a correction to the
Schr\"odinger equation.  
If this is done in the way GRW propose,
the predictions for experiments carried out on microscopic 
systems are almost precisely the same, so that none of the
successes of quantum theory in this realm are lost.  
However, large systems deviate more significantly from the predictions
of quantum theory.
Those deviations are still quite subtle, and very hard to 
detect or exclude experimentally at present, but they 
are unambiguously there in the equations.  Experimentalists will 
one day be able to tell us for sure whether or not they are there 
in nature.  

By making this modification, we turn quantum theory into a theory
which describes objective events continually taking place in 
a real external world, whether or not any experiment is 
taking place, whether or not anyone is 
watching.  If this picture is right, it solves the measurement
problem: we
have a single set of equations which give a unified description of
microscopic and macroscopic physics, and we can 
sensibly talk about the behaviour of unobserved systems, whether
they are microscopic electrons or macroscopic pointers.  The pointer
of an apparatus probing a quantum system takes up a definite position,
and does so very quickly, not through any ad hoc postulate, but in
a way that follows directly from the fundamental equations of the
theory.  

The GRW theory is probably completely wrong in detail.  
There are certainly serious difficulties in making it 
compatible with relativity --- though there also some grounds for 
optimism that this can be done (Pearle 1998, Kent 1999).
But GRW's essential idea has, I think, a fair chance of
being right.  Before 1986, few people believed that any tinkering
with quantum theory was possible: it seemed that any change must
so completely alter the structure of the theory as to violate
some already tested prediction.  But we now know that 
it is possible to make relatively tiny changes which cause no 
conflict with experiment, and that by doing so we can solve
the deep conceptual and interpretational problems of quantum
theory.  We know too that the modified theory 
makes new experimental predictions in an entirely unexpected 
physical regime.   The crucial tests, if and when we can carry
them out, will be made not by probing deeper into the nucleus or 
by building higher energy accelerators, but by keeping relatively
large systems under careful enough control for quantum effects
to be observable.  
New physics could come directly from the large scale and the complex: 
frontiers we thought long ago closed.  

\newsec{Physics and Consciousness}

Kieslowski's remarkable film series, Dekalog,
begins with the story of a computer scientist and his son who
share a joy in calculating and predicting, in using the computer
to give some small measure of additional control over their lives. 
Before going skating, the son obtains weather reports for the 
last three days from the meteorological bureau, and together they run
a program to infer the thickness of the ice and deduce that 
it can easily bear his weight.  But, tragically, they neglect 
the fire a homeless man keeps burning at the lakeside.     
Literally, of course, they make a simple mistake: the 
right calculation would have taken account of the fire,
corrected the local temperature, and shown the actual
thickness of the ice.  Metaphorically, the story seems to say
that the error is neglecting the spiritual, not only in life, 
but perhaps even in physical predictions.  

I do not myself share Kieslowski's religious worldview, and
I certainly do not mean to start a religious discussion here. 
But there is an underlying scientific question, which can
be motivated without referring to pre-scientific systems of
belief and is crucial to our understanding 
of the world and our place in it, and which I think is 
still surprisingly neglected.  So, to use more scientifically 
respectable language, I would like to take a fresh look 
at the problem of consciousness in physics, where by ``consciousness''
I mean the perceptions, sensations, thoughts 
and emotions that constitute our experience.   

There has been a significant revival of interest in consciousness lately,
but it still receives relatively little attention from physicists.
Most physicists believe that, 
if consciousness poses any problems at all,
they are problems outside their province.\foot{Penrose is the
best-known exception: space does not permit discussion of his
rather different arguments here, but see Penrose 1989, 1994.}
After all, the argument runs, biology is pretty much
reducible to chemistry, which is reducible to known physical
laws.  Nothing in our current understanding suggests that
there is anything physically distinctive about living beings,
or brains.  On the contrary, neurophysiology, experimental psychology,
evolutionary and molecular biology have all advanced
with great success, based firmly on the hypothesis that there
is not.  Of course, no one can exclude the possibility that our current
understanding could turn out to be wrong --- but in the
absence of any reason to think so, there seems nothing useful 
for physicists to say.  

I largely agree with this view.  
It {\it is} very hard to see how any novel physics
associated with consciousness could fit with what we
already know.  Speculating about such ideas {\it does}
seem fruitless in the absence of data. 
But I think we can say something.  There is
a basic point about the connection between consciousness
and physics which ought to be made, yet
seems never to have been clearly stated, and which suggests our
present understanding almost cannot be complete.  

The argument for this goes in three steps. 
First, let us assume, as physicists quite commonly do,
that any natural phenomenon can be described mathematically.  
Consciousness is a natural phenomenon, and at least
some aspects of consciousness --- for example, the number of 
symbols we can simultaneously keep in mind --- 
are quantifiable.   On the other hand we have no 
mathematical theory even of these aspects of consciousness.  
This would not matter if we could at least sketch a path
by which statements about consciousness could be reduced
to well understood phenomena.  After all, no one worries
that we have no mathematical theory of digestion, because
we believe that we understand in principle how to rewrite
any physical statement concerning the digestive process as
a statement about the local densities of various
chemicals in the digestive tract, and how to derive
these statements from the known laws of physics.
But we cannot sketch a similar path for consciousness:
no one knows how to transcribe a statement of the 
form ``I see a red giraffe'' into a statement about
the physical state of the speaker.  
To make such a transcription, we would need to attach a theory of 
consciousness to the laws of physics we know: it clearly 
cannot be derived from those laws alone. 

Second, we note that, despite the lack of a theory of consciousness,
we cannot completely keep consciousness out of physics.  
All the data on which our theories are based ultimately 
derive from  
conscious impressions or conscious memories of impressions.
If our ideas about physics included no hypothesis about 
consciousness, we would have no way of deriving any 
conclusion about the data, and so no logical reason for preferring
any theory over any other. 
This difficulty has long been recognised.  It 
is dealt with, as best we can, by  
invoking what is usually called the principle of
psycho-physical parallelism.  We demand that
we should at least be able to give 
a plausible sketch of how an accurate representation of the contents 
of our conscious minds could be included in the description of the
material world provided by our physical theories, assuming 
a detailed understanding of how consciousness is represented. 

Since we do not actually know how to represent consciousness, 
that may seem an empty requirement, but it is not.  
Psycho-physical parallelism requires, for example, that a theory
explain how anything that we may observe can come to be 
correlated with something happening in our brains, and
that enough is happening in our brains at any given moment
to represent the full richness of our conscious experience.
These are hard criteria to make precise,
but asking whether they could plausibly be satisfied within
a given theory is still a useful constraint.

Now the principle of psycho-physical parallelism,
as currently applied, commits us to seeing 
consciousness as an epiphenomenon supervening on the
material world.  As William James magnificently put
it (James 1879): 

{\narrower\smallskip\noindent Feeling is a mere collateral product of
our nervous processes, unable to react upon them any
more than a shadow reacts on the steps of the traveller
whom it accompanies.  Inert, uninfluential, a simple passenger
in the voyage of life, it is allowed to remain on board, but
not to touch the helm or handle the rigging.\smallskip} 

Third, the problem with all of this is, 
that as James went on to point out, if 
our consciousness is the result
of Darwinian evolution, as it surely must be, 
it is difficult to understand how it can be an epiphenomenon.  
To sharpen James' point: if there is a simple mathematical theory
of consciousness, or of any quantifiable aspect of
consciousness, describing a precise version of the principle
of psycho-physical parallelism and so characterising how 
it is epiphenomenally attached
to the material world, then its apparent evolutionary
value is fictitious.   For all the difference it would
make to our actions, we might as well be 
conscious only of the number of neutrons in our kneecaps or the
charm count of our cerebella;
we might as well find pleasures painful and vice versa.    
In fact, of course, our consciousness tends to supply us
with a sort of executive summary of information with a direct bearing
on our own chances of survival and those of our genes; we tend
to find actions pleasurable or painful depending whether
they are beneficial or harmful to those chances. 
Though we are not always aware of vital information, and are
always aware of much else, and though our preferences certainly
don't perfectly correlate with our genetic prospects, the 
general predisposition of consciousness towards survival
is far too strong to be simply a matter of chance.  

Now, of course, almost no one seriously suggests that the main
features of consciousness can be the way they are purely by chance.
The natural hypothesis is that, since they seem to be evolutionarily
advantageous, they should, like our other evolutionarily advantageous
traits, have arisen through a process of natural selection.
But if consciousness really is an epiphenomenon, this explanation
cannot work.  An executive summary of information which is presented 
to us, but has no subsequent influence on our behaviour, carries
no evolutionary advantage.  It may well be advantageous for us 
that our brains run some sort of higher-level processes which use
the sort of data that consciousness presents to us and which are 
used to make high-level decisions about behaviour.  But, on the
epiphenomenal hypothesis, we gain nothing by being conscious of
these particular processes: if they are going to run, they
could equally well be run unconsciously, leaving our attention
focussed on quite different brain activities or on none at all.

Something, then, is wrong with our current understanding, 
There are really only two serious possibilities.  
One is that psycho-physical parallelism cannot be made precise and that
consciousness is simply scientifically inexplicable.   
The other is that consciousness is something which
interacts, if perhaps very subtly, with the rest of the material 
world rather than simply passively co-existing alongside that world.  
If that were the case, then we can think of our consciousnesses
and our brains --- more precisely, the components of our brains
described by presently understood physics --- as two coupled systems, 
each of which influences the other.  That is a radically different
picture from the one we presently have, of course.  But it does have
explanatory power.  If it were true,
it would be easy to understand why it might be evolutionarily
advantageous for our consciousness to take a particular form.
If say, being conscious of a particular feature of the environment
helps to speed up the brain's analysis of that feature, or to
focus more of the brain's processing power on it, or to execute
relevant decisions more quickly, or to cause a more sophisticated
and detailed description to enter into memory, then evolution 
would certainly cause consciousness to pay attention to the relevant  
and neglect the irrelevant.  

We have to be clear about this, though: to propose this explanation
is to propose that the actions of conscious beings
are not properly described by the present laws of physics.
This does not
imply that conscious actions cannot be described by any laws.
Far from it: if {\it that} were the case, we would still have 
an insoluble mystery, and 
once we are committed to accepting an insoluble mystery associated
with consciousness then we have no good reason to prefer a mystery
which requires amending the laws of physics over one which leaves
the existing laws unchallenged.  The scientifically interesting
possibility --- the possibility with maximal explanatory power --- 
is that our actions and those of other conscious beings are not
perfectly described by the laws we presently know, but could be 
by future laws which include a proper theory
of consciousness.  

This need not be true, of course.  Perhaps
consciousness {\it will} forever be a mystery.  But it seems
hard to confidently justify any {\it a priori} division
of the unsolved problems in physics into the soluble and the
forever insoluble.  We ought at 
least to consider the implications of maximal ambition.  
We generally assume that everything
in nature except consciousness has a complete mathematical 
description: that is why, for example, we carry on looking for a way of
unifying quantum theory and gravity, despite the apparent
difficulty of the problem.  We should accept that, if
this assumption is right, it is at least plausible that consciousness
also has such a description.  And this forces us to accept
the corollary --- that there is a respectable case for 
believing that we will eventually find we need new 
dynamical laws --- even though nothing else we
know supports it.  

One final comment: nothing in this argument relies on
the peculiar properties of quantum theory, or the problems
it poses.  The argument runs through equally well in
Newtonian physics.  Maybe the deep problems of quantum theory
and consciousness are linked, but it seems to me we have
no reason to think so.  It follows that anyone committed 
to the view I have just outlined must argue that a deep problem
in physics has generally been neglected for the last century and a half.
So let me try to make that case.  

There is no stronger or more venerable scientific taboo
than that against enquiry, however tentative, into consciousness.  
James, in 1879, quoted ``a most intelligent biologist'' as saying:

{\narrower\smallskip\noindent It is high time for scientific
men to protest against the recognition of any such thing as
consciousness in scientific investigation. \smallskip} 

Scientific men and women certainly have protested this, loudly and often,
over the last hundred and twenty years.
But have those protests ever carried much intellectual force?

The folk wisdom, such as it is, against the possibility of
a scientific investigation of 
consciousness seems now to rest on a confusion 
hanging over from the largely deleterious effect of logical
positivism on scientists earlier this century.  Hypotheses
about consciousness are widely taken to be {\it ipso facto}
unscientific because consciousness is presently unmeasurable
and its influences, if any, are presently undetectable.  
Delete the word ``presently'', and the case could be properly
made: as it is, it falls flat.
If logical positivism is to blame, is only 
the most recent recruit to the cause.  The problem seems to run much
deeper in scientific culture.  Schr\"odinger described (Schr\"odinger
1954) the phenomenon of:

{\narrower\smallskip\noindent [$\ldots$] the wall, separating the `two paths',
that of the heart and that of pure reason.  We look back along the
wall: could we not pull it down, has it always been there?  As we
scan its windings over hills and vales back in history we behold
a land far, far, away at a space of over two thousand years back,
where the wall flattens and disappears and the path was not yet split,
but was only {\it one}.  Some of us deem it worth while to walk back
and see what can be learnt from the alluring primeval
unity.

Dropping the metaphor, it is my opinion that the philosophy of the
ancient Greeks attracts us at this moment, because never before or
since, anywhere in the world, has anything like their highly
advanced and articulated system of knowledge and speculation
been established {\it without} the fateful division which has
hampered us for centuries and has become unendurable in our days.
\smallskip}

Clearly, the revival of interest in Greek philosophy that Schr\"odinger saw
did not immediately produce the revolution he hoped for. 
But our continued fascination with consciousness
is evident on the popular science and philosophy bookshelves.
It looks as though breaking down the wall and building 
a complete worldview are going to be left
as tasks for the third millennium.  There could hardly be 
greater or more fascinating challenges.  

Nor can there be many more necessary for our long term well being.  Science 
has done us far more good than harm, psychologically and 
materially.  But the great advances we have made in understanding nature
have also been used to support a worldview in which only
what we can now measure matters, in which the material 
and the external dominate, in which we objectify and
reduce ourselves and each other, in which we are in danger
of coming to see our psyches and our cultures, in all
their richness, as no more than the evolutionarily honed
expression of an agglomeration of crude competitive urges.

To put it more succinctly, there is a danger, as
Vaclav Havel put it in a recent essay (Havel 1996), of 
man as an observer
becoming completely alienated from himself as a being. 
Havel goes on to suggest that hopeful signs of a more humane and
less schizophrenic worldview can be found in what he suggests
might be called postmodern science, in the form of the Gaia
hypothesis and the anthropic principle.  

I disagree: it is hard to pin down precise scientific 
content in these ideas, and insofar as we can 
it seems to me they are no help.  But I think we have the answer
already.  The alienation is an artefact, created by the erroneous
belief that all that physics currently describes is all there is. 
But, on everything we value in our humanity, physics is silent.
As far as our understanding of human consciousness is concerned,
though we have learned far more about ourselves, we have 
learned nothing for sure that negates or delegitimizes a 
humane perspective.  In that sense, nothing of crucial 
importance has changed.   

\newsec{Postscript} 

All this said, of course, predicting the future of science is a  
mug's game.  If, as I have argued, physics is very far from over,
the one thing we should be surest of is that greater surprises
than anything we can imagine are in store.  
One prediction that seems likelier than most, though, is that
the Editor will not be restricted   
to considering human contributors for the corresponding 
volume in 2999.  Perhaps our future extraterrestrial or mechanical
colleagues will find some amusement in our attempts.  I do hope so.      

\vskip.3in
\leftline{\bf References}\par
\vskip.1in
\item{}
Schr\"odinger, E.~1954  {\it Nature and the Greeks}.  
Cambridge: Cambridge University Press.
\vskip.1in
\item{} Bell, J.S.~1987.  {\it Speakable and Unspeakable in Quantum
Mechanics: Collected papers on Quantum Philosophy}. 
Cambridge: Cambridge University Press
\vskip.1in
\item{} Feynman, R.~1965  {\it The Character of Physical Law}. 
London: British Broadcasting Corporation.
Reading: Addison Wesley.
\vskip.1in
\item{} Snow, C.P.~1934  {\it The Search}.  London: Victor Gollancz.
\vskip.1in
\item{} Ghirardi, G.~et al.~1986  Unified Dynamics for Microscopic and
Macroscopic Systems.  {\it Physical Review D} {\bf 34} 470-491.
\vskip.1in
\item{} Landau, L.~and Lifshitz, E.~1974  {\it Quantum Mechanics}.
Oxford: Pergamon Press.
\vskip.1in
\item{} Pearle, P.~1999  Relativistic Collapse Model with Tachyonic
Features.  {\it Physical Review A} {\bf 59} 80-101.
\vskip.1in
\item{} Kent, A.~1998  Quantum Histories.  {\it Physica Scripta} 
{\bf T76} 78-84.
\vskip.1in
\item{} Penrose, R.~1989  {\it The Emperor's New Mind: Concerning
Computers, Minds, and the Laws of Physics}.
Oxford: Oxford University Press.
\vskip.1in
\item{} Penrose, R.~1994  {\it Shadows of the Mind: A Search for the
Missing Science of Consciousness}.
Oxford: Oxford University Press.
\vskip.1in
\item{}  James, W.~1879  Are We Automata?   {\it Mind} {\bf 13} 1-22.
\vskip.1in
\item{}
Havel,V.~ 1996.  In {\it The Fontana Postmodernism Reader}, (ed.
W. Truett Anderson).   London: Fontana.
\vskip.1in
\end